In this part you will find the LateX source. In parts 2 and 3 you
will find Postscript files for pages 20-23, and 24-25.

\documentstyle [12pt] {article}

\newcommand{\be}{\begin{equation}}
\newcommand{\ee}{\end{equation}}
\newcommand{\bes}{\begin{eqnarray}}
\newcommand{\ees}{\end{eqnarray}}
\newcommand{\tr}{{\rm Tr}}
\newcommand{\retr}{{\rm Re}\,\tr}
\newcommand{\R}{{\vec{R}}}
\begin{document}
\begin{titlepage}
\rightline{WUB 92--29}
\rightline{August 1992}
\vskip 2cm
\begin{center}
\Large
{\bf Running Coupling and the\\
$\Lambda$-Parameter from \\
SU(3) Lattice Simulations}\footnote{Work
supported by EC project SC1*--CT91--0642,
DFG grant Schi 257/1--4}
\end{center}
\vskip 1.6cm
\centerline{\bf Gunnar S.~Bali\footnote{E-mail:
bali@wrcm1.urz.uni-wuppertal.de}  and  Klaus~Schilling}
\centerline{Physics Department, Bergische Universit\"at,
             Gesamthochschule Wuppertal}
\centerline{Gau\ss{}stra\ss{}e 20, 5600 Wuppertal, Germany}
\vskip 1.5cm
\centerline{\bf ABSTRACT}
\sl
\small
We present new results on the static $q\bar q$ potential
from high statistics simulations
on $32^4$ and smaller lattices, using
the standard Wilson action at $\beta = 6.0, 6.4$,
and $6.8$ on the Connection Machine CM-2. Within our statistical
errors ($\approx 1\%$) we do not observe any finite size effects
affecting the potential values, on varying the spatial lattice extent from
$0.9f\!m$ up to $3.3f\!m$. We are able to see and
quantify the running of the coupling from the Coulomb behaviour
of the interquark force. From this we extract the ratio
$\sqrt{\sigma}/\Lambda_L$.
We demonstrate that scaling violations on the string tension
can be considerably  reduced by
introducing effective coupling schemes, which allow for a safe
extrapolation of $\Lambda_L$ to its continuum value.
Both methods yield consistent values
for $\Lambda$:
$\Lambda_{\overline{MS}}=0.558_{-0.007}^{+0.017}\times\sqrt{\sigma}
=246_{-3}^{+7}MeV$.
At the highest energy scale attainable to us we find
$\alpha(5\,GeV)=0.150(3)$.\vskip .3cm
\noindent\small PACS numbers: 11.15.Ha, 12.38.Gc, 12.38.Aw
\vfill\eject
\end{titlepage}

\vspace*{3 true cm}

\normalsize
\section{Introduction}
The experimental determination of the running coupling {\em constant} of
QCD has reached a reasonable
degree of accuracy~\cite{hebekker} after two decades of research
effort.
This has stimulated considerable attention to compute this quantity
from first principles, by use of lattice
methods~\cite{mack,chris,su2schroed}.
The lattice approach to the problem of matching perturbative
and nonperturbative aspects of QCD
is notoriously difficult because of the requirement of a high energy
resolution.
Nevertheless, computer experiments in
pure $SU(2)$ and $SU(3)$ gauge theory have reached a precision
that allows to ask rather detailed questions about
the static quark-antiquark potential.
The size of the available lattices ($48^3\times 56$,
in $SU(2)$ gauge theory~\cite{UKQCD})
enables one
to decrease the lattice spacing $a$ into a regime
where one can make contact to predictions of continuum
perturbation theory.
This has been done for the case of $SU(2)$
by a study of  the Coulomb behaviour of the
interquark force in ref.~\cite{chris}.
In the case of $SU(3)$,
a lattice spacing of $a^{-1} = 3.6 \,GeV$
was achieved so far~\cite{wir} on a $32^4$ lattice at $\beta = 6.4$.
This resolution is about the treshold for  running coupling effects
to become visible.

In this paper we want to
present a detailed investigation of the running coupling in $SU(3)$
gauge theory,
by further reducing the lattice spacing to $a^{-1} = 6.0 \,GeV$.
Within our analysis of the small distance
regime, we will use a parameterization incorporating
lattice effects.
Being limited to  lattice sizes up to $32^4$, we have to
make sure that our results are not spoiled by finite size
effects. For this reason we have worked on a variety of
lattices, at each value of $\beta$.

Once the running coupling has been extracted, we
will be able to compare to perturbative predictions and
estimate  a value for the corresponding $\Lambda_L$ parameter.
We will see that this
value is consistent  with $\Lambda_L$,
as obtained   from the string tension
(by the use of the two-loop $\beta$-function~\cite{Hasenfratz}),
after an extrapolation to $a = 0$.
In order to substantiate this result,
we will improve on scaling
violations (as expressed
in the strong $\beta$-dependence of $\Lambda_L$) by  replacing
the bare coupling with suitable  ``effective''
couplings~\cite{parisi,par2,par3,fing}, measured
on the lattice from the average plaquette.
In this case, we will find nearly  asymptotic scaling
for $\beta> 6.0$.
The extrapolation to the continuum
yields an estimate for $\Lambda_L$ which is consistent within smaller
errors with the value obtained from the running coupling.

\section  {Methods}
\subsection {Sampling}
In order to maintain an appropriate stochastic movement
of the gauge system through
phase space with increasing $\beta$,
we have combined
one Cabibbo-Marinari pseudo-heatbath-sweep~\cite{CAB} over the three
diagonal $SU(2)$ subgroups with
{\it four(nine)} successive  overrelaxation
sweeps~\cite{Adler} for $\beta = 6.4 (6.8)$.
We reach an acceptance rate of $99.5 \%$ for an overrelaxation link update.
For the heatbath we use the algorithm proposed by Kennedy and
Pendleton~\cite{Pendleton} which has a high acceptance rate and can
thus
be efficiently implemented on a SIMD machine. We can afford
iterating the algorithm
until all link variables are changed. On our local 8K CM-2 we need
$9.2 \mu$sec for an overrelaxation link update and $11.5 \mu$sec for a single
Cabibbo-Marinari link update. This performance was achieved after
rewriting the $SU(3)$ matrix multiply routines in assembler language.
Measurements were started after 2000 -- 10000 thermalization sweeps.

\subsection {Smoothing Operators}
In lattice gauge theory physical quantities of interest like masses,
potential values, and matrix elements are related to asymptotic properties
of exponentially decreasing correlation functions in Euclidean time, and
therefore prone to be drowned in noise. So one is forced to improve
operators in order to reach the desired asymptotic behaviour for
the small $T$ region. We will shortly describe our particular
improvement technique~\cite{wir}.

We start from the relation between Wilson loops, $W(R,T)$, and the (ground
state) potential $V(R)$
\be
W(R,T) = C(R)\exp\left\{-TV(R)\right\}\quad +\quad
{\rm excited\,\, state\,\, contributions.}
\ee
Our aim is to enhance --- for each value of $R$ --- the corresponding
ground state overlap $C(R)$. Since the ground state wave function is
expected to be smooth on an ultraviolet scale we concentrate on reducing
noise by applying a {\it local} smoothing procedure on the spatial links:
consider a spatial link variable $U_i(n)$, and the sum of the four spatial
staples $\Pi_i(n)$ connected to it:
\be
\Pi_i(n) = \sum_{\stackrel{j = \pm 1,\ldots,3}{^{j\neq i}}}
             U_j(n)U_i(n+\hat{j})
             U_j^{\dagger}(n+\hat{i}).
\ee
We apply a gauge covariant, iterative smoothing algorithm which replaces
(in the same even/odd ordering as the Metropolis update) $U_i(n)$ by
$U'_i(n)$ minimizing the local spatial action
$S_i(n) = -\retr\{U_i(n)\Pi^{\dagger}_i(n)\}$, which is qualitatively
a measure for the {\it roughness} of the gauge field. This is very similar to
lattice cooling techniques already invented by previous
authors~\cite{HOEK,PISA} except that we are {\it cooling} only within
time-slices and thus not affecting the transfer matrix. Alternatively, this
algorithm may be interpreted as substituting
$U_i(n)$ by ${\cal P}\left(\Pi_i(n)\right)$ where ${\cal P}$ denotes
the projection operator onto the {\it nearest} $SU(3)$ matrix.
In this sense it is a variant of the APE recursive
blocking scheme~\cite{APE} with
the coefficient of the straight link set to {\it zero}, but with even/odd
updating. The latter feature renders the algorithm less memory consuming
and seems to improve convergence. Contributions
from excited states become increasingly suppressed, as we repeat this
procedure. After
$30 (45)$ such smoothing steps at $\beta = 6.4 (6.8)$ we reach values for
the overlap $C(R)$ of $95(80)\%$ for small (large) spatial separations $R$.

\subsection {Extraction of Potential Values}
For the extraction of the potential from the Wilson loop data we
proceed essentially as described in ref.~\cite{wir}, with a slight
modification that helps to carry out a straightforward error analysis.
Instead of fitting the Wilson loops to the dependence
\be
\label{eq-fit}
W_{\R}(T;C(\R),V(\R)):= C(\R)\exp\left\{-V(\R)T\right\}
\ee
for $T\geq T_{min}$ with some reasonable cutoff $T_{min}$
we take the {\em local mass}
\be
\label{eq-locmass}
V_{T_{min}}(\R) = \ln\left\{\frac{W(\R,T_{min})}{W(\R,T_{min}+1)}\right\}
\ee
as an estimator for the potential $V(\R)$.
By using this explicit formula for the calculation of $V(\R)$,
we are able to propagate the covariance matrix between Wilson loops to a
covariance matrix for the potential values.
This allows one to separate the determination of potential parameters
from the measurement of the potential itself, helping to decrease the
degrees of freedom and promoting stability within the fitting procedure.
Note that the value of $V(\R)=V_{T_{min}}(\R)$ does not differ appreciably
from the result of a fit to eq.~(\ref{eq-fit})
because the latter is anyhow dominated by the
lowest two $T$ data due to their small relative errors.

The optimization of the overlap $C(\R)$ proceeds as described
in ref.~\cite{wir}:
The parameters $C(\R)$ and $V(\R)$ are fitted
for different $T_{min}$ to the Wilson loop
data separately for each smoothing step (and $\R$) according to
eq.~(\ref{eq-fit}) by minimizing
\bes
\chi_{\R}^2(C(\R),V(\R)) =
\sum_{T1,T2}&\left(W(\R,T_1)-W_{\R}(T_1;C(\R),V(\R))\right)
\left(C^{\R\R}\right)^{-1}_{T_1T_2}\\\nonumber
\times &\left(W(\R,T_2)-W_{\R}(T_2;C(\R),V(\R))\right).
\ees
$C^{\R_1\R_2}_{T_1T_2}$ denotes the covariance matrix which is
estimated to be
\be
C^{\R_1\R_2}_{T_1T_2}=\frac{1}{N(N-1)}\sum_{i=1}^N
\left(W_i(\R_1,T_1)-W(\R_1,T_1)\right)
            \left(W_i(\R_2,T_2)-W(\R_2,T_2)\right).
\ee
We have divided the timeseries of Wilson loops into $N$ successive subsets of
given length $n$. $W_i(\R,T)$ stands for the average of the respective
Wilson loop over the $i$th subset. $n$ should be chosen such that
$\tau\ll n\ll N$, in order to cope with the autocorrelation time $\tau$.
Afterwards for each value of $\R$ the smoothing
step with highest ground state overlap $C(\R)$ is selected from the fits
with reasonable $\chi^2$.

In a second step stability of local masses $V_T(\R)$ (eq.~(\ref{eq-locmass}))
against variation of $T$ is checked, and $T_{min}(\R)$ is determined
as the $T$ value (plus {\it one}) from which onwards stability within
errors is observed. For large $R$ values we find
$T_{min} = 4$. For simplicity we chose the same value for small $R$.

As promised, we are now able to propagate the
covariance matrix between different
Wilson loops $C_{T_1T_2}^{\R_1\R_2}$ to a covariance matrix between
the potential values $C_V^{\R_1\R_2}$, by using the quadratic
approximation\footnote{
In order to check the validity of this approximation, we have
moreover carried out a bootstrap
analysis~\cite{bootstrap} of our
data on the $32^4$ lattices. (This method is
also shortly described in the appendix of ref.~\cite{rajan}.)
The resulting errors (and biased values) are almost identical
with the results of our approximation, but
the bootstrap method alone does not deliver reliable $\chi^2$ values
(incorporating the correlation effects).}
\bes
C^{\R_1\R_2}_V &=& \sum_{T_1T_2}
    \frac{\partial V(\R_1)}{\partial W(\R_1,T_1)}
    C_{T_1T_2}^{\R_1\R_2}\frac{\partial V(\R_2)}{\partial W(\R_2,T_2)}\\
&=&\frac{C_{T(\R_1),T(\R_2)}^{\R_1\R_2}}
        {W(\R_1,T(\R_1))W(\R_2,T(\R_2))}
  +\frac{C_{T(\R_1)+1,T(\R_2)+1}^{\R_1\R_2}}
        {W(\R_1,T(\R_1)+1)W(\R_2,T(\R_2)+1)}\nonumber\\
&-&\frac{C_{T(\R_1)+1,T(\R_2)}^{\R_1\R_2}}
        {W(\R_1,T(\R_1)+1)W(\R_2,T(\R_2))}
  +\frac{C_{T(\R_1),T(\R_2)+1}^{\R_1\R_2}}
        {W(\R_1,T(\R_1))W(\R_2,T(\R_2)+1)}\nonumber
\ees
where $T(\R)$ is used as an abbreviation for $T_{min}(\R)$.
With this covariance matrix we are able to fit the potential data to
various parameterizations, incorporating all possible correlations
between different operators measured on individual configurations as
well as correlation effects within the Monte Carlo time series of
configurations.

\subsection{Measurements}
The lattice parameters used for the simulations\footnote{Note
that we have adapted the physical scales from
$\sqrt{\sigma}=420\,MeV$ to $\sqrt{\sigma} = 440\,MeV$.} are collected
in table~\ref{tab-lattice} which includes
quotations of $32^4$
lattices at $\beta = 6.0$, and $\beta = 6.4$, as well as a
$24^3\times 32$ lattice at $\beta = 6.4$ that have been simulated
recently~\cite{wir}, and are reanalysed in the present investigation.
The spatial extent of the lattices at $\beta=6.4$ ranges from
$aL_S=0.87\,f\!m$ to $1.74\,f\!m$. At $\beta = 6.8$
lattice volumes of
$(0.52\,f\!m)^3$ and $(1.05\,f\!m)^3$ have been realised.
The resolution $a^{-1}$ is varied from $1.9\,GeV$ to $6.0\,GeV$.

Smoothened on- and off-axis Wilson loops were measured every 100
sweeps (every 50 sweeps for $\beta = 6.0$). Up to
$N_{max} = 30(45)$ smoothing steps were performed
at $\beta=6.0, 6.4(6.8)$.
The following spatial separations were realized: $\R = M\vec{e}_i$
with $\vec{e}_i = (1,0,0),(1,1,0),(2,1,0),(1,1,1),(2,1,1),(2,2,1)$.
$M$ was increased up to $L_S/2$ for $i = 1,2,4$, and up to
$L_S/4$ for the remaining directions. Altogether this yields 72
different separations $\R$ on the $32^3\times L_T$ lattices. The
time separations $T=1,2,\ldots,10$ were used. Thus the total
number of operators measured on one configuration ($V = 32^3\times L_T$)
is $72\times 10\times N_{max}$.

\begin{table}
\caption[The simulated lattices]{The simulated lattices. Physical units
correspond to the choice
\(\sqrt{\sigma} = 440 MeV\) for the string tension.
Errors ignore the experimental uncertainty within the value
of the string tension.}
\label{tab-lattice}

{\footnotesize \begin{center}\begin{tabular}{|c||r||r|r|r|r||r|r|}\hline
&$\beta=6.0$&\multicolumn{4}{c||}{$\beta = 6.4$}
&\multicolumn{2}{c|}{$\beta = 6.8$} \\ \hline
$V = L_S^3\times L_T$
&$32^4$&$16^3\times 32$&$24^3\times 32$&$32^3\times 16$&$32^4$&$16^3\times
64$&$32^4$\\ \hline
$a/f\!m$
&0.101 (2)
&\multicolumn{4}{c||}{0.0544 (5)}&\multicolumn{2}{c|}{0.0327 (5)}\\ \hline
$a^{-1}/GeV$
&1.94 (5)&\multicolumn{4}{c||}{3.62 (4)}&\multicolumn{2}{c|}{6.02 (10)}\\
\hline
$aL_S/f\!m$
&3.25 (8)&0.87 (1)&1.31 (1)&\multicolumn{2}{c||}{1.74 (2)}&0.52 (1)&1.05 (2)\\
\hline
$(aL_T)^{-1}/MeV$
&61 (1)&\multicolumn{2}{c|}{113 (1)}&226 (2)&113 (1)&94 (2)&188 (3)\\\hline
Total \# of sweeps&6100&11900&22000&10000&8900&20400&15900\\ \hline
Thermalization phase&1000&2000&2100&2100&2500&10000&5000\\ \hline
\# of measurements&102&100&200&80&65&105&110\\ \hline
\end{tabular}\end{center}}
\end{table}

The potential values at $\beta=6.0$, and $\beta=6.4$ have been listed in our
previous publication~\cite{wir}. For convenience of the reader we
collect the corresponding values for $\beta=6.8$ in the appendix.

\section {Results}
\subsection{$q\bar q$-Potential}

\begin{table}
\caption[Results of the $q\bar q$ potential fits]{Fit results.
Since the parameter values on the largest
lattices are most precise, we refrain from citing results gained on
smaller volumes as long as they are compatible with the stated numbers.
For the $16^3\times 64$ lattice at $\beta = 6.8$ this is not the
case. Therefore we have listed both
the standard fit result, and the parameter values with the
string tension constrained to its $32^4$ value.\\\vspace{.3cm}}
\label{fitres}

{\small \begin{center}\begin{tabular}{|c|c|c|c|c|c|}\hline
&$\beta=6.0$&$\beta=6.4$&\multicolumn{3}{c|}{$\beta=6.8$}\\\hline
Vol.&$32^4$&$32^4$&$32^4$&\multicolumn{2}{c|}{$16^3\times 64$}\\ \hline
$K$&0.0513 (25)&0.01475 (29)&0.00533 (18)&0.00545(27)&0.00533\\ \hline
$e$&0.275 (28)&0.315 (15)&0.311 (10)&0.269 (22)&0.274 (18)\\ \hline
$V_0$&0.636 (10)&0.6013 (37)&0.5485 (24)&0.5412 (37)&0.5426 (34)\\ \hline
$l$&0.64 (12)&0.564 (55)&0.558 (35)&0.725 (87)&0.710 (120)\\ \hline
$f$&0.041 (58)&0.075 (18)&0.094 (13)&0.037 (26)&0.043 (25)\\ \hline
$R_{\rm min}$&2&$\sqrt{3}$&$\sqrt{3}$&$\sqrt{3}$&$\sqrt{3}$\\ \hline
$\frac{\chi^2}{N_{DF}}$&0.816&0.953&0.937&0.989&0.754\\\hline
\end{tabular}\end{center}}
\end{table}

We connect our investigation to the recent $SU(2)$ analysis by Chris
Michael~\cite{chris}, and start from his ansatz:
\be
\label{eq-pot}
V(\R)=V_0+KR-e\left(\frac{1-l}{R} + l\,4\pi
G_L(\R)\right)+\frac{f}{R^2}.
\ee
The lattice propagator for the one gluon exchange~\cite{Rebbi}
\be
\label{eq-glue}
G_L(\R)=\int_{-\pi}^{\pi}\frac{d^3k}{(2\pi)^3}\frac{\cos(\vec{k}\R)}{4\sum_i
\sin^2(k_i/2)}.
\ee
has been calculated numerically. The parameter $l$ is expected to be
in the range $0\leq l \leq 1$ and controls the violation of rotational
symmetry on the lattice (within this ansatz). The term $f/R^2$ mocks
deviations from a pure Coulomb behaviour and is expected to be positive
to the extent that asymptotic freedom becomes visible in the effective
Coulomb term $-(e-f/R)/R$.

A test of the ansatz eq.~(\ref{eq-pot}) implies that the ``corrected'' data
$V(R)=V(\R)+\delta V(\R)$ with
\be
\label{eq-deltaV}
\delta V(\R) = el\left(4\pi G_L(\R) - 1/R\right)
\ee
are independent of the direction of $\R$. The global situation
is depicted  for the
$32^4$ lattice at $\beta = 6.4$ in figure~1 where the corrected data
points are plotted together with the interpolating fit
$V(R) = V_0 + K R-e/R+f/R^2$, with fit parameters $V_0,K,e$, and $f$
as given in table~\ref{fitres}.
Our potential fits yield $\chi^2/N_{DF}<1$
as long as the first two\footnote{Three for $\beta = 6.0$.} data points are
excluded. The stability of
the string tension result with respect to cuts in $R$ is displayed in
figure~2 (for $\beta=6.4$, and $6.8$).

For $\beta\geq 6.4$ the Coulomb coefficients $e$ are
definitely different from the value $\pi/12\approx 0.262$ predicted by
the string vibrating picture~\cite{Luescher} for
large $q\bar q$ separations. The self energy contribution $V_0$
follows the leading order expectation $V_0\propto 1/\beta$.
We emphasize that for all $\beta$ values the parameter $f$ is established to
be positive as expected. In fact, this parameter tends to increase
with $\beta$, weakening the Coulomb coupling for small distances.

A more sensitive representation of the scatter of the data points
around the interpolating fit curve (obtained on the $32^4$ lattice)
is shown in figure~3 (for $\beta=6.4$). Note that the deviations
are within a $1\%$ band for the
largest volume, once the first two data points are excluded.
Decreasing the lattice spatially or in the time direction by a factor
of two leaves the data points compatible with the interpolating curve,
i.e.\ the finite size effects (FSE) are below our statistical accuracy.
Nevertheless it pays to work on a $32^4$ lattice since the
larger possible $q\bar q$ separations increase
the lever arm needed to fix the long distance part of the potential.

At $\beta=6.8$ we find indications of FSE by comparing
results from the small lattice and the $32^4$ lattice.
As the string tension appears not to
suffer from these effects, we have fixed its value
to that measured on the larger lattice in order to study FSE
on the remaining parameters more directly. The largest FSE
occurs for the lattice correction parameter $l$. This may be due to
the low momentum cutoff that starts to become visible on the
scale of a few lattice spacings. By chosing the form of the one gluon exchange
(eq.~(\ref{eq-glue})), we have neglected this cutoff in the integral
bounds.

We concentrate our interest here on short distance physics where
the linear term is not yet dominating the potential. In the case
of $\beta=6.4$ the latter happens at $R\approx 5$.
{}From figure~3 we conclude that reliable results can be extracted
for a lattice as small as $16^3$ for this $\beta$-value. In physical
units this corresponds to a $27^3$ lattice at $\beta=6.8$. So a
volume of $32^3$ (or even smaller) appears to be sufficiently large for
our purpose.

A synopsis of data for $\beta=6.0, 6.4$, and $6.8$, in physical units,
is displayed in figure~4 with logarithmic ordinate ranging from
$0.03\,f\!m$
up to $1.9\,f\!m$. The three data sets collapse to a universal potential.
The two curves correspond to a linear plus
Coulomb parameterization, with the string tension
$\sigma=Ka^{-2}=(440 MeV)^2$, and
the strength of the Coulomb term determined by our fit to the
$\beta=6.4$ data ($e=0.315$, full curve), and fixed to the L\"uscher-value
($e = \pi/12$, dashed curve), respectively. The plot demonstrates the
incompatibility of the data points with a pure Coulomb behaviour for
short distances, and the necessity of additional terms like $f/R^2$.

\subsection{Running Coupling}
Our lattice analysis for the running coupling $\alpha_{q\bar q}(R)$
closely follows
the procedure suggested in ref.~\cite{chris}. We start from the
symmetric discretization in terms of the force $F$
\be
\label{alpha}
\alpha_{q\bar q}(R)=-\frac{3}{4}R_1R_2 F(R)
=\frac{3}{4}R_1R_2\frac{V(R_1)-V(R_2)}{R_1-R_2}.
\ee
with $R=(R_1+R_2)/2$.
We take the corrected potential $V(R_i)=V(\R_i)+\delta V(\R_i)$ with $\delta
V(\R_i)$ as given in eq.~(\ref{eq-deltaV}).
Unlike ref.~\cite{chris}, however, we use all possible combinations $\R_1,\R_2$
with $|\R_1-\R_2| < 1.5$.

The resulting data points are contained in figure~5a. In order to exhibit
both the global behaviour, and the perturbative region ($R\rightarrow
0$) we decided to use a logarithmic ordinate (in units of
$\sigma^{-1/2}$). The latter region is expanded in the inset.
We omitted all values with errors
$\Delta\alpha_{q\bar q}(R)>\alpha_{q\bar q}(R)/3$ in order not to
clutter the graph.
In addition to the statistical
error of the force $F(R)$ we allow for a systematic error
\be
\Delta F_{syst}(R)
            = \left(\left(\frac{\Delta l}{l}\right)^2
               + \left(\frac{\Delta e}{e}\right)^2\right)^{1/2}
              |\delta F(R)|
\ee
with
$\delta F(R) =  \frac{\delta V(\R_1)-\delta V(\R_2)}{R_1-R_2}$.
$\Delta F_{syst}$ is typically of the order of $10 \%$ of the lattice
correction $\delta F(R)$.

Now we can proceed to analyse our $\alpha_{q\bar q}$-data in terms of the
continuum large momentum
expectation for the running coupling:
\be
\label{alpha2}
\alpha_{q\bar q}(R)=\frac{1}{4\pi}\left(b_0\ln\left(Ra\Lambda_R\right)^{-2}
        +b_1/b_0\ln\ln\left(Ra\Lambda_R\right)^{-2}\right)^{-1},
\ee
with
\be
b_0=\frac{11}{3}\frac{N_C}{16\pi^2},\quad
b_1=\frac{34}{3}\left(\frac{N_C}{16\pi^2}\right)^2
\ee
being the first two coefficients of the weak coupling expansion of
the $SU(N_C)$ Callan-Symanzik $\beta$-function
(eq.~(\ref{beta}) below).
In order to extract $\Lambda_R$ we base our fits exclusively on data points
at $\beta=6.8$ with $R_1,R_2\geq\sqrt{3}$ on the r.h.s.\ of
eq.~(\ref{alpha}). This is done in order to avoid the danger of
``pollution'' from
discretization errors.

We now ask the question, within which $R$ region our data are
compatible --- if at all! --- with the asymptotic behaviour of
eq.~(\ref{alpha2}). We find that as long as $R\sqrt{K}<0.173$ our fits
yield results with reasonable $\chi^2/N_{DF}$. This upper limit in $R$
corresponds to $2.5\,GeV$. Fitting the $\beta=6.8$ data over this
region we obtain
\be
\Lambda_R=(0.562\pm 0.020\pm 0.010)\sqrt{\sigma}\approx (247\pm 10) MeV.
\ee
The first error stems from the fit just described, while the second
relates to
the statistical uncertainty of the string tension within our lattice analysis.
The corresponding fit curve with error bands is plotted in
figure~5.
As the data appear to osculate the asymptotic curve one finds a
systematic dependence on the $R$ cut:
$\Lambda_R$ tends to be larger if more (low energy)
data points are included and
{\em vice versa}. In this sense one might consider our value as an
upper limit.

Exploiting the relation $\Lambda_R = 30.19\Lambda_L$~\cite{billoire}
we get:
\be
\Lambda^{\alpha}_L=(18.6\pm 0.7\pm 0.3)
\times 10^{-3}\sqrt{\sigma}\approx (8.19\pm 0.33) MeV.
\ee
This corresponds to the ratio
\be
\label{abc}
\frac{\sqrt{\sigma}}{\Lambda_L^{\alpha}}=53.7\pm 2.1.
\ee
In figure~5b we have plotted $\alpha$ versus the energy. At
the largest realized energy scale we find
$\alpha_{q\bar q}(5\,GeV)\approx 0.150(3)$.

Returning to the global structure of the data displayed in figure~5
we make
three observations: 1.~The small $R$ contributions (circles, and
triangles) follow very neatly the asymptotic perturbative prediction
eq.~(\ref{alpha2}), indicating very little discretization effects.
2.~Over the whole $R$
range the data sets for $\beta=6.4$, and $\beta=6.8$ coincide very
nicely, giving evidence for scaling. 3.~The deviations of the data from
the asymptotic behaviour remain
fairly small up to $q\approx 1 GeV$ or
$\alpha_{q\bar q}\approx 0.4$.

We conclude that lattice simulations can indeed make contact to the
perturbative regime. Moreover, it is very satisfying to observe that
the 2-loop-formula describes the lattice data down to a scale as small
as $1\,GeV$
--- at least in the quenched approximation of QCD. One would expect that
the situation in full QCD is fairly similar, concerning this property.
In the infrared regime ($q<\sqrt{\sigma}$)
the differences between both theories will be considerable:
Because of the linear confining potential our expectation for
the pure gauge sector is $\alpha_{q\bar q}(q)\propto 1/q^2$. This has
to be confronted with the
expression $\alpha_{q\bar q}(q)\propto
e^{-\mu/q}/q^2$ for QCD with fermionic degrees of freedom where $\mu$
stands for the screening mass.

\subsection{Scaling}

\begin{table}
\caption[The lattice spacing $a$, and the
parameter $\Lambda_L$ as a function of $\beta$]{The lattice spacing $a$,
and cutoff parameters $\Lambda_L$
calculated from the 2-loop-expansion eq.~(\ref{twoloo}) in units of
the string tension $\sigma$. $\Lambda_L$ is obtained by inserting the
bare lattice coupling. For $\Lambda_L^{(1,2)}$ the $\beta_E^{(1,2)}$
effective couplings were used. A naive linear
extrapolation to $a=0$ leads to the results displayed in the second
last row. Logarithmic extrapolations yield the values in the last row.
\\\vspace{.3cm}}
\label{tab-string}

{\begin{center}\begin{tabular}{|c|c|c|c|c|c|}\hline
\multicolumn{2}{|c|}{$\beta$}&$a\sqrt{\sigma}$&$\sqrt{\sigma}/\Lambda_L$&
$\sqrt{\sigma}/\Lambda_L^{(1)}$&$\sqrt{\sigma}/\Lambda_L^{(2)}$\\\hline
\multicolumn{2}{|c|}{5.7}&0.4099 (24)&124.7 (0.7)&63.3 (0.4)&55.7 (0.3)\\\hline
\multicolumn{2}{|c|}{5.8}&0.3302 (30)&112.4 (1.0)&63.0 (0.6)&55.6 (0.5)\\\hline
\multicolumn{2}{|c|}{5.9}&0.2702 (37)&102.9 (1.4)&61.2 (0.8)&54.3 (0.7)\\\hline
\multicolumn{2}{|c|}{6.0}&0.2265 (55)& 96.5 (2.3)&60.0 (1.5)&53.4 (1.3)\\\hline
\multicolumn{2}{|c|}{6.2}&0.1619 (19)& 86.4 (1.0)&56.9 (0.7)&50.8 (0.6)\\\hline
\multicolumn{2}{|c|}{6.4}&0.1215 (12)& 81.3 (0.8)&55.7 (0.5)&50.0 (0.5)\\\hline
\multicolumn{2}{|c|}{6.8}&0.0730 (12)& 76.9 (1.3)&55.7 (0.9)&50.4
(0.8)\\\hline\hline
$\infty$&lin.&0& 63.6 (2.4)&53.1 (1.6)&48.3 (1.4)\\\cline{2-6}
        &log.&0&$54^{+18}_{-15}$&$53.2^{+2.6}_{-7.3}$
&$49.1^{+2.3}_{-5.9}$\\\hline
\end{tabular}\end{center}}
\end{table}

\begin{table}
\caption[The average plaquette action $\langle S_{\Box}\rangle$]{The
average plaquette action $\langle S_{\Box}\rangle$,
measured on large lattice volumes. The values for $\beta\leq 5.9$ are
taken from the collection in ref.~\cite{fing} while the other numbers
are our new results, obtained on $32^4$
lattices, and one $24^3\times 32$ lattice ($\beta=6.2$).}
\label{tab-pla}

{\begin{center}\begin{tabular}{|c|c|}\hline
$\beta$&$\langle S_{\Box}\rangle$\\\hline
5.7&0.45100 (80)\\\hline
5.8&0.43236  (5)\\\hline
5.9&0.41825  (6)\\\hline
6.0&0.406262(17)\\\hline
6.2&0.386353 (8)\\\hline
6.4&0.369353 (5)\\\hline
6.8&0.340782 (5)\\\hline
\end{tabular}\end{center}}
\end{table}

Normally one
speaks of asymptotic scaling when the ratio $\sqrt{\sigma}/\Lambda_L$
remains constant on varying $\beta$ where
\be
\label{twoloo}
\Lambda_L= \frac{1}{a}
\exp\left(-\frac{1}{2b_0g^2}\right)(b_0g^2)^{-\frac{b_1}{2b_0^2}}
\ee
(with $g^2=2N_C/\beta$)
denotes the integrated two-loop $\beta$-function (eq.~(\ref{beta}) below).
In table~\ref{tab-string} we have compiled
our new results on the string tension
together with previous results from refs.~\cite{wir,MTC}.
As can be Seen, we are still far away from the asymptotic scaling
region up to $\beta=6.8$.

We attempt to extrapolate $\Lambda_L^{-1}$ to the continuum limit by
the use of a
parameterization that takes into account the leading order expectation
for scaling violations ${\cal O}(1/\ln a)$:
\be
\label{extra}
\Lambda_L^{-1}(a)=\Lambda_L^{-1}(0)
+\frac{C}{\sqrt{\sigma}\ln (Da\sqrt{\sigma})}.
\ee
We find the data compatible with this logarithmic behaviour, with
$D\approx 1$--$2$, and $C\approx 20$--$80$.
The fit parameters are not particularly stable with respect to a
variation of the number of data points. The bandwith of
extrapolations to the continuum limit is illustrated in figure~6a
where we have plotted the extreme cases of a fit to our four low $a$
data points, and all seven data points (open circles). If we average the
values obtained from these fits, and take the upmost
and the lowest possible numbers as error bandwidth, we estimate
the asymptotic value to be
$\sqrt{\sigma}\Lambda_L(0)^{-1}=54^{+18}_{-15}$ (full circle).
We would like to mention that a
naive linear extrapolation to the continuum limit yields the value
$\sqrt{\sigma}\Lambda_L(0)^{-1}=63.6 (2.4)$ with (obviously)
underestimated error. We take this
as a warning for purely phenomenological continuum extrapolations.

In view of the uncertainty of the above number
it would be highly desirable to improve the situation by developing a
scheme within which the $a$ dependence of $\Lambda_L(a)$ is reduced.
Parisi suggested many years ago a more ``natural'' expansion
parameter $g_E$~\cite{parisi}, based on a mean field argument.
His scheme was elaborated in refs.~\cite{par2,par3,fing}.
It works as follows:
Let $c_n$ be the coefficients of the weak coupling expansion of the
average plaquette
\bes
\label{weak}
\langle S_{\Box}\rangle
&=&\frac{1}{6V}\sum_{\Box}
   \left(1-\frac{1}{N_C}\retr\, U_{\Box} \right)\nonumber\\
&=&\sum_{n=1}^{\infty}c_ng^{2n}.
\ees
The idea, now, is to introduce an effective coupling in terms of
the Monte Carlo generated average plaquette
\bes
g_E^2&=&\frac{\langle S_{\Box}\rangle}{c_1}\\
&=&g^2+\frac{c_2}{c_1}g^4+\frac{c_3}{c_1}g^6+{\cal O}(g^8),\nonumber
\ees
for which the first order expansion is exact.
The hope is that the
nonperturbative (or higher order perturbative) contributions that are
resummed in the effective coupling $g_E$ may compensate high order
terms in the $\beta$-function which are responsible for the
scaling violations.
Support for this expectation comes from the observed
scaling of ratios of physical quantities (figures~4,~5) within the
same $\beta$ region.

The coefficients $c_1$, and $c_2$ have been calculated
previously~\cite{digiac}, and an unpublished value for $c_3$, obtained
by H.~Panagopoulos, has been cited in ref.~\cite{fing}. The numerical
values are:
\bes
c_1&=&(N_C^2-1)/(8N_C)\\
c_2&=&(N_C^2-1)(0.0204277-1/(32N_C^2))/4\\
c_3&=&(N_C^2-1)N_C(0.0066599-0.020411/N_C^2+0.0343399/N_C^4)/6.
\ees
The plaquette values needed for the conversion into the effective
coupling schemes are collected in table~\ref{tab-pla}.
The numbers for $\beta\leq
5.9$ were taken from the collection in ref.~\cite{fing}.
Starting from the expansion
\be
\label{beta}
\beta(g)=-\frac{dg}{d\ln\!a}=-\sum_{n=0}^{\infty}b_ng^{2n+3}
\ee
of the $\beta$-function, one rewrites
\bes
\label{xxx}
\beta(g_E)
&=&-\frac{dg_E}{d\ln\!a}=-\frac{dg}{d\ln\!a}\frac{g}{g_E}\frac{dg_E^2}{dg^2}\\
&=&-b_0g_E^3-b_1g_E^5-b_2g_E^7+
\left(3b_0\left(2\left(\frac{c_2}{c_1}\right)^2
-\frac{c_3}{c_1}\right)-2b_1\frac{c_2}{c_1}\right)g_E^7+{\cal
O}(g_E^9).\nonumber
\ees

The first two terms in this weak coupling expansion remain unchanged
under the substitution.
Therefore, an integration again leads to
eq.~(\ref{twoloo}), but with a redefined integration constant
\be
\Lambda_E=\Lambda_L\exp\left(\frac{c_2}{2c_1b_0}\right)
\approx 2.0756\Lambda_L\quad({\rm for}\quad SU(3)).
\ee
This factor is due to a shift of the effective $\beta$ by a constant
in the continuum limit: $g_E^{-2}=g^{-2}-c_2/c_1+{\cal O}(g^2)$.
In the following we will refer to this scheme as the
$\beta_E^{(1)}$ scheme.
As one can see from figure~6a (open squares), and table~\ref{tab-string}
this kind of (numerical) resummation of
the asymptotic series eq.~(\ref{weak}) leads to considerably reduced
logarithmic corrections ($C\approx 2.5$).

As an additional check of this improvement technique we consider
in the following
an ``alternative'' effective coupling scheme $\beta_E^{(2)}$. Our
idea is to introduce a coupling $g_{2}$ by inverting the relation
\be
\langle S_{\Box}\rangle = c_1g_{2}^2+c_2g_{2}^4.
\ee
This amounts to truncating the weak coupling expansion
eq.~(\ref{weak}) after the {\em second} term\footnote{One can
generalise this scheme by truncating in higher orders $n$.
This is of little interest, however (unless one is interested in
numerical studies of the impact of a particular higher loop
contribution on the observed scaling violations),
since the $\beta$-function has only
been calculated up to ${\cal O}(g^5)$. Moreover, one would retrieve the
bare coupling scheme at $n$ sufficiently large.}.
A short calculation yields:
\be
\label{yyy}
\beta(g_{2})=-b_0g_{2}^3-b_1g_{2}^5-b_2g_{2}^7
-3b_0\frac{c_3}{c_1}g_{2}^7
+{\cal O}(g_{2}^9).
\ee
Because of
$g_{2}^{-2}=g^{-2}+{\cal O}(g^2)$
the integration constant $\Lambda_L$ remains unchanged in respect to
the original bare coupling scheme.

If we compare the third order terms of the two effective schemes
(eqs.~(\ref{xxx},\ref{yyy})), we
find explicitely:
\be
\beta(g)=\beta(g_{E}(g))+5.3\times 10^{-4}g^7 +{\cal O}(g^9)
=\beta(g_{2}(g))+4.02\times10^{-3}g^7+{\cal O}(g^9).
\ee
This means that the correction  of the
$\beta$-function through the 3-loop-contribution
is much larger for the $\beta^{(2)}$ than for
the $\beta^{(1)}$ scheme\footnote{Note that the difference between the
$\beta$-functions for both effective
schemes is independent of $c_3$ to this order.}.
Nevertheless, at least within the investigated $\beta$ region,
the qualitative behaviour of both schemes is the same as can be seen
in figure~6a. For the
$\beta^{(2)}$ scheme the correction coefficient
($C\approx 0.9$) of the continuum extrapolation
eq.~(\ref{extra})
is even smaller than for the $\beta_E^{(1)}$ scheme. In figure~6a we
have included the estimates for the asymptotic $\Lambda_L^{-1}$ values
(and the $\Lambda_L^{-1}$ from the running coupling) as full symbols.

The extrapolated values for both effective schemes are, respectively:
\bes
\sqrt{\sigma}&=&53.2^{+2.6}_{-7.3}\Lambda_L^{(1)}\\
             &=&49.1^{+2.3}_{-5.9}\Lambda_L^{(2)}.
\ees
Averaging these numbers that carry asymmetric errors leads to
$\sqrt{\sigma}=50.8^{+1.0}_{-4.6}\Lambda_L^E$.
This result is in nice agreement with the ratio extracted from
the running coupling
($\sqrt{\sigma}=53.7(2.1)\Lambda_L^{\alpha}$, eq.~(\ref{abc})).
Using this additional information,
we obtain:
\be
\sqrt{\sigma}=51.6^{+0.7}_{-1.6}\Lambda_L.
\ee
This result may be converted into any continuum
renormalization scheme like the minimal subtraction ($\overline{MS}$)
scheme. By exploiting the relation
$\Lambda_{\overline{MS}}=28.81\Lambda_L$~\cite{msbar},
we get:
\be
\frac{\Lambda_{\overline{MS}}}{\sqrt{\sigma}}=0.558_{-0.007}^{+0.017}.
\ee

Let us finally comment that the two approaches presented in this paper
for the determination of the QCD scale parameter $\Lambda$,
namely to analyze (a) $g^2(\Lambda aR)$, and (b)
its inverse $\Lambda a(g^2)$ in
terms of the two-loop predictions eqs.~(\ref{alpha2},\ref{twoloo}),
are complementary and supportive to each other
because higher order corrections to methods (a) and (b) are
anticorrelated.
In our running coupling (string tension) analysis we observe the
``effective'' $\Lambda_L$ to decrease (increase) with the energy
scale. Since the central value of our ``upper limit''
$\Lambda_L^{\alpha}$ is smaller than that of our ``lower limit''
$\Lambda_L^E$ we are in the position to state relatively small errors for
$\Lambda_{\overline{MS}}$.

In figure~6b we have plotted the $\Lambda_L^{-1}$ data versus $\beta$
in order to visualize the slow approach of the bare coupling data
towards the asymptotic value, and the improvement achieved
by the use of {\em effective} couplings.

\section{Discussion}

We have demonstrated that medium size computer experiments are able to
determine the $\Lambda$-parameter of $SU(3)$ Yang-Mills theory within
a reasonable accuracy (that can compete with QCD experiments).
For this result, it has been important to study both infrared,
and ultraviolett aspects in order to verify the reliability of the
continuum extrapolation. We might say that we have been lucky to get
hold of {\em asymptotia} within our means.
This is due to the discovery that the
running coupling {\em constant} is well described within this theory
by the two-loop formula down to a scale of about $1\, GeV$.

If nature continues to be nice to us, and the inclusion of dynamical
quarks results only in a $\beta$-shift of quenched predictions
it is possible to
predict experimental numbers like $\alpha_S(M_Z)$, as explained in
ref.~\cite{mack}. Obviously, it is preferable to repeat this
study in full QCD on the level of TERAFLOPS power. In the meantime,
further improvements of lattice techniques are
of great interest. A promising route has been proposed by M.~L\"uscher
{\em et.~al.}~\cite{schroed}, and tested on
$SU(2)$ Yang-Mills theory. These authors start from a volume dependent
coupling $g(L)$ which allows them to reach large energies on small
lattices.

After completion of this work we received a preprint by S.P.~Booth,
C.~Michael, and collaborators~\cite{runsu3} that contains a running
coupling study for $SU(3)$ gauge theory up to $\beta=6.5$. Their
results are fully consistent with ours.

\vspace*{.5cm}

\noindent{\bf Acknowledgements.} We are grateful to Deutsche
Forschungsgemeinschaft for the support given to our CM-2 project.
We thank Peer Ueberholz and Randy Flesch
for their kind support. One of the authors (G.B.) would like to thank
Chris Michael, Edwin Laermann, Rainer Sommer, and Jochen Fingberg
for helpful discussions about data analysis, and the
different effective coupling schemes.
\vfill\eject

\noindent {\LARGE\bf Appendix}
\begin{appendix}

\section{Potential Values}
\label{app2}

In this appendix we are stating the potential values measured on a
$32^4$ lattice at $\beta=6.8$. The corresponding numbers for the other
$\beta$-values can be found in ref.~\cite{wir}. The on- and
off-axis paths are numbered in the following way:

\begin{center}\begin{tabular}{|c|c|c|c|c|c|c|}\hline
Path \#&1&2&3&4&5&6\\\hline
Path $(X,Y,Z)$&$(1,0,0)$&$(1,1,0)$&$(2,1,0)$&$(1,1,1)$&$(2,1,1)$&
               $(2,2,1)$\\\hline
Elementary distance $M$& $1$&$1.41$&$2.24$&$1.73$&$2.45$&$3$\\\hline
\end{tabular}\end{center}

The results for the potential $V(\R)$ (in lattice units), as well
as for the ``corrected'' $V(R)$, and
the corresponding ground state overlaps $C(\R)$ are
collected in table~5. The data is plotted (among the other
curves) in figure~4.

\noindent Table~5: The potential values $V(\R)$ (in lattice units $a^{-1}$),
``corrected'' values $V(R)$, and groundstate overlaps
$C(\R)$ for $\beta = 6.8$, $V = 32^4$.
\begin{center}
\begin{tabular}{|c|c|c|c|c|}\hline
$R$&Path&$V(\R)$&$V(R)$&$C(\R)$\\\hline
 1.00&1&0.3107  (6)&0.3210 (10)&0.950  (3)\\\hline
 1.41&2&0.3855 (11)&0.3794 (12)&0.951  (4)\\\hline
 1.73&4&0.4188 (19)&0.4098 (20)&0.946  (8)\\\hline
 2.00&1&0.4236 (14)&0.4266 (14)&0.929  (5)\\\hline
 2.24&3&0.4428 (13)&0.4397 (14)&0.934  (5)\\\hline
 2.45&5&0.4559 (15)&0.4509 (15)&0.936  (6)\\\hline
 2.83&2&0.4696 (20)&0.4656 (20)&0.923  (8)\\\hline
 3.00&1&0.4725 (14)&0.4709 (14)&0.931  (6)\\\hline
 3.00&6&0.4751 (18)&0.4705 (19)&0.924  (7)\\\hline
 3.46&4&0.4906 (31)&0.4861 (31)&0.923 (12)\\\hline
 4.00&1&0.5000 (18)&0.4970 (19)&0.916  (7)\\\hline
 4.24&2&0.5079 (23)&0.5039 (23)&0.939  (9)\\\hline
 4.47&3&0.5105 (22)&0.5068 (22)&0.916  (9)\\\hline
\end{tabular}\end{center}
\vfill\eject

\begin{center}Table~5, continued\\

\begin{tabular}{|c|c|c|c|c|}\hline
$R$&Path&$V(\R)$&$V(R)$&$C(\R)$\\\hline
 4.90&5&0.5178 (28)&0.5139 (28)&0.913 (11)\\\hline
 5.00&1&0.5193 (19)&0.5159 (20)&0.924  (8)\\\hline
 5.20&4&0.5230 (32)&0.5190 (32)&0.929 (13)\\\hline
 5.66&2&0.5312 (29)&0.5273 (30)&0.920 (12)\\\hline
 6.00&1&0.5325 (25)&0.5289 (25)&0.907 (10)\\\hline
 6.00&6&0.5357 (29)&0.5317 (30)&0.918 (12)\\\hline
 6.71&3&0.5421 (27)&0.5383 (27)&0.917 (11)\\\hline
 6.93&4&0.5469 (42)&0.5430 (42)&0.916 (16)\\\hline
 7.00&1&0.5463 (27)&0.5426 (27)&0.921 (11)\\\hline
 7.07&2&0.5474 (36)&0.5436 (37)&0.928 (15)\\\hline
 7.35&5&0.5504 (32)&0.5466 (32)&0.923 (13)\\\hline
 8.00&1&0.5568 (34)&0.5531 (34)&0.910 (13)\\\hline
 8.49&2&0.5623 (44)&0.5584 (44)&0.911 (17)\\\hline
 8.66&4&0.5644 (47)&0.5605 (47)&0.930 (19)\\\hline
 8.94&3&0.5663 (37)&0.5625 (37)&0.911 (15)\\\hline
 9.00&1&0.5671 (36)&0.5633 (36)&0.920 (14)\\\hline
 9.00&6&0.5651 (37)&0.5612 (37)&0.911 (15)\\\hline
 9.80&5&0.5733 (41)&0.5695 (41)&0.908 (16)\\\hline
 9.90&2&0.5745 (48)&0.5707 (48)&0.925 (19)\\\hline
10.00&1&0.5743 (44)&0.5705 (44)&0.904 (17)\\\hline
10.39&4&0.5777 (53)&0.5739 (53)&0.905 (21)\\\hline
11.00&1&0.5830 (49)&0.5792 (49)&0.913 (19)\\\hline
11.18&3&0.5818 (49)&0.5780 (49)&0.903 (20)\\\hline
11.31&2&0.5841 (50)&0.5803 (50)&0.898 (20)\\\hline
12.00&1&0.5887 (55)&0.5849 (55)&0.895 (21)\\\hline
12.00&6&0.5900 (55)&0.5862 (55)&0.901 (22)\\\hline
12.12&4&0.5941 (60)&0.5902 (60)&0.928 (24)\\\hline
12.25&5&0.5918 (53)&0.5879 (53)&0.912 (21)\\\hline
12.73&2&0.5962 (64)&0.5923 (64)&0.918 (26)\\\hline
13.00&1&0.5987 (56)&0.5949 (56)&0.912 (22)\\\hline
\end{tabular}\end{center}
\vfill\eject

\begin{center}Table~5, continued\\

\begin{tabular}{|c|c|c|c|c|}\hline
$R$&Path&$V(\R)$&$V(R)$&$C(\R)$\\\hline
13.42&3&0.5998 (61)&0.5960 (61)&0.894 (24)\\\hline
13.86&4&0.6031 (75)&0.5993 (75)&0.895 (29)\\\hline
14.00&1&0.6055 (62)&0.6017 (62)&0.899 (24)\\\hline
14.14&2&0.6052 (73)&0.6014 (73)&0.893 (29)\\\hline
14.70&5&0.6096 (70)&0.6058 (70)&0.895 (27)\\\hline
15.00&1&0.6097 (68)&0.6059 (68)&0.895 (27)\\\hline
15.00&6&0.6102 (69)&0.6064 (69)&0.895 (27)\\\hline
15.56&2&0.6139 (81)&0.6101 (81)&0.904 (32)\\\hline
15.59&4&0.6163 (81)&0.6125 (81)&0.910 (32)\\\hline
15.65&3&0.6144 (73)&0.6106 (73)&0.895 (29)\\\hline
16.00&1&0.6151 (74)&0.6113 (74)&0.878 (29)\\\hline
16.97&2&0.6246 (88)&0.6209 (88)&0.886 (34)\\\hline
17.15&5&0.6248 (78)&0.6210 (78)&0.895 (31)\\\hline
17.32&4&0.6258 (94)&0.6220 (94)&0.883 (36)\\\hline
17.89&3&0.6296 (90)&0.6258 (90)&0.880 (35)\\\hline
18.00&6&0.6312 (88)&0.6274 (88)&0.885 (34)\\\hline
18.39&2&0.6337 (99)&0.6299 (99)&0.899 (39)\\\hline
19.05&4&0.6394(109)&0.6357(109)&0.900 (43)\\\hline
19.60&5&0.6402 (95)&0.6364 (95)&0.874 (36)\\\hline
19.80&2&0.6440(105)&0.6402(105)&0.882 (41)\\\hline
20.79&4&0.6486(117)&0.6448(117)&0.875 (44)\\\hline
21.00&6&0.6496(108)&0.6458(108)&0.879 (42)\\\hline
21.21&2&0.6526(118)&0.6489(118)&0.891 (46)\\\hline
22.52&4&0.6610(131)&0.6573(131)&0.889 (51)\\\hline
22.63&2&0.6545(128)&0.6508(128)&0.847 (48)\\\hline
24.00&6&0.6688(123)&0.6650(123)&0.863 (47)\\\hline
24.25&4&0.6694(142)&0.6657(142)&0.862 (53)\\\hline
25.98&4&0.6791(151)&0.6753(151)&0.866 (57)\\\hline
27.71&4&0.6908(162)&0.6871(162)&0.848 (60)\\\hline
\end{tabular}\end{center}
\vfill\eject

\end{appendix}

\vfill\eject

\listoftables
\contentsline {table}{\numberline {5}{\ignorespaces The potential values for
$\beta =6.8$, $V=32^4$}}{14}
\vspace{3cm}
\listoffigures
\contentsline {figure}{\numberline {1}{\ignorespaces Static $q\mathaccent "7016
q$ potential for $\beta = 6.4$}}{20}
\contentsline {figure}{\numberline {2}{\ignorespaces Impact of low $R$-cuts on
the string tension}}{21}
\contentsline {figure}{\numberline {3}{\ignorespaces Finite Volume effects at
$\beta = 6.4$}}{22}
\contentsline {figure}{\numberline {4}{\ignorespaces Scaling plot of potential
data obtained for $\beta =6.0$, 6.4, and 6.8}}{23}
\contentsline {figure}{\numberline {5}{\ignorespaces The running coupling}}{24}
\contentsline {figure}{\numberline {6}{\ignorespaces Test on asymptotic scaling
of the string tension}}{25}


\begin{thebibliography}{99}

\bibitem{hebekker} {See e.g.~the review article: T.~Hebbeker, ``Tests
of QCD\ldots'', to appear in {\em Phys.~Reports}.}

\bibitem{mack} {A.X.~El-Khadra, G.~Hockney, A.S.~Kronfeld, and
P.B.~Mackenzie, Fermilab preprint 91/354-T; P.B.~Mackenzie,
{\em Nucl.~Phys.}\ {\bf B}[Proc.~Suppl.]{\bf 26} (1992) 369.}

\bibitem{chris} {C.~Michael, {\em Phys.~Lett.}\ {\bf B283} (1992) 103.}

\bibitem{su2schroed}{M.~L\"uscher, R.~Sommer, U.~Wolff, and P.~Weisz,
CERN preprint CERN-TH 6566/92.}

\bibitem{UKQCD} {The UKQCD Collaboration: S.P.~Booth, K.C.~Bowler,
D.S.~Henty, R.D.~Kenway, B.J.~Pendleton,
D.G.~Richards, A.D.~Simpson, A.C.~Irving, A.~McKerrell, C.~Michael,
P.W.~Stephenson, M.~Teper, and K.~Decker, {\em Phys.~Lett.}\
{\bf B275} (1992) 424.}

\bibitem{wir} {G.S.~Bali and K.~Schilling, Wuppertal preprint WUB
92--02, to appear in {\em Phys.~Rev.} {\bf D46} (1992).}

\bibitem{Hasenfratz} {A.~and P.~Hasenfratz, {\em Phys.~Lett.}\
                 {\bf 93B} (1980) 165;\\
                  {\em Nucl.~Phys.}\ {\bf B193} (1981) 210.}

\bibitem{parisi} {G.~Parisi, Proceedings of the {\sc xx}th
International Conference on High Energy Physics 1980, Madison, Eds.\
L.~Durand, and L.G.~Pondrom, American Institute of Physics, New York
(1981) 1531.}

\bibitem{par2}{Y.M.~Makeenko and M.I.~Polikarpov, {\em Nucl.~Phys.}\
              {\bf B205} (1982) 386.}

\bibitem{par3}{S.~Samuel, O.~Martin, and K.~Moriarty, {\em Phys.~Lett.}\
              {\bf 152B} (1984) 87.}

\bibitem{fing}{J.~Fingberg, U.~Heller, and F.~Karsch, Bielefeld
preprint BI-TP 92-26}

\bibitem{CAB} {N.~Cabibbo and E.~Marinari, {\em Phys.~Lett.}\
                            {\bf 119B} (1982) 387.}

\bibitem{Adler} {S.L.~Adler, {\em Phys.~Rev.}\ {\bf D23} (1981) 2901.}

\bibitem{Pendleton} {A.~Kennedy and B.~Pendleton, {\em Phys.~Lett.}\
                     {\bf B156} (1985) 393.}

\bibitem{HOEK} {J.~Hoek, M.~Teper, and J.~Waterhouse,
               {\em Nucl.~Phys.}\ {\bf B288} (1987) 589.}

\bibitem{PISA} {M.~Campostrini, A.~Di~Giacomo, M.~Maggiore,
                H.~Panagopoulos, and E.~Vicari, {\em Phys.~Lett.}\ {\bf B225}
                (1989) 403.}

\bibitem{APE}  {The APE Collaboration: M.~Albanese {\em et al.},
                {\em Phys.~Lett.}\ {\bf B192} (1987) 163.}

\bibitem{bootstrap} {B.~Efron, {\em Ann.~Statist.} {\bf 7} (1979) 1.}

\bibitem{rajan} {R.~Gupta {\em et.~al.}, {\em Phys.~Rev.}\
               {\bf D36} (1987) 2813.}

\bibitem{Rebbi} {C.B.~Lang and C.~Rebbi, {\em Phys.~Lett.}\
                      {\bf 115B} (1982)137.}

\bibitem{Luescher} {M.~L\"uscher, K.~Symanzik, and P.~Weisz, {\em Nucl.~Phys.}\
                   {\bf B173} (1980) 365; M.~L\"uscher, {\em Nucl.~Phys.}\
                   {\bf B180} (1981) 317.}

\bibitem{billoire} {A.~Billoire, {\em Phys.~Lett.}\ {\bf 104B} (1981)
                    472.}

\bibitem{MTC} {The $MT_C$ Collaboration: K.D.~Born, R.~Altmeyer,
               W.~Ibes, E.~Laermann, R.~Sommer,
               T.F.~Walsh, and P.~Zerwas,
               {\em Nucl.~Phys.}\ {\bf B}[Proc.~Suppl.]{\bf 20} (1991)
               394.}

\bibitem{digiac} {A.~DiGiacomo and G.C.~Rossi, {\em Phys.~Lett.}\
                 {\bf 100B} (1981) 481;
                  A.~DiGiacomo and G.~Paffati, {\em Phys.~Lett.}\
                 {\bf 108B} (1982) 327;
                  U.~Heller and F.~Karsch, {\em Nucl.~Phys.}
                 {\bf B251} (1985) 254.}

\bibitem{msbar}{R.~Dashen and D.J.~Gross, {\em Phys.~Rev.}\
               {\bf D23} (1981) 2340.}

\bibitem{schroed} {M.~L\"uscher, R.~Narayanan, P.~Weisz, and U.~Wolff,
DESY preprint 92-025, to appear in {\em Nucl.~Phys.}\ {\bf B}.}

\bibitem{runsu3}{S.P.~Booth, D.S.~Henty, A.~Hulsebos, A.C.~Irving,
C.~Michael, and P.W.~Stephenson, Liverpool preprint LTH 285 (1992).}

\end{thebibliography}
\end{document}